\documentclass[aip,jcp,reprint,twocolumn,superscriptaddress]{revtex4-1}

\usepackage{amsmath,mathtools}
\usepackage{amssymb}
\usepackage{graphicx}
  \usepackage{hyperref}
\usepackage[arrow, matrix, curve]{xy}
\usepackage{bm}
\usepackage{color}

\newcommand{\bra}[1]{\mbox{$\langle #1 |$}}

\newcommand{\ket}[1]{\mbox{$| #1 \rangle$}}

\DeclareMathAlphabet\mathbfcal{OMS}{cmsy}{b}{n}

\usepackage[usenames,dvipsnames]{xcolor}
\hypersetup{colorlinks=true, linkcolor=BrickRed, urlcolor=blue!50!black, citecolor=blue!50!black}

\usepackage[normalem]{ulem}

\makeatletter
\newcommand\hide@visible[1]{%
  \bgroup\fboxsep=.3ex\colorbox{Gray}{begin hide}%
  #1\colorbox{Gray}{end hide}\egroup%
}
\newcommand\hide@hidden[1]{%
  \bgroup\fboxsep=.3ex\colorbox{Gray}{hidden text}%
}
\newcommand\hide@invisible[1]{}
\newcommand\makevisible{\let\hide\hide@visible}
\newcommand\makehidden{\let\hide\hide@hidden}
\newcommand\makeinvisible{\let\hide\hide@invisible}
\makeatother

\usepackage{amsmath}

\makehidden

\begin{document}
\title{Relating the pure and ensemble density matrix functional}

\author{Christian Schilling}
\email{christian.schilling@physics.ox.ac.uk}
\affiliation{Clarendon Laboratory, University of Oxford, Parks Road, Oxford OX1 3PU, United Kingdom}

\date{\today}


\begin{abstract}
A crucial theorem in Reduced Density Matrix Functional Theory (RDMFT) suggests that the universal pure and ensemble functional
coincide on their common domain of pure $N$-representable one-matrices.
We refute this by a comprehensive analysis of the geometric picture underlying Levy's constrained search.
Moreover, we then show that the ensemble functional follows instead as the lower convex envelop of the pure functional. It is particularly remarkable that the pure functional determines the ensemble functional even outside its own domain of pure $N$-representable one-matrices.
From a general perspective, this demonstrates that relaxing pure RDMFT to ensemble RDMFT does not necessarily circumvent the complexity of the one-body pure $N$-representability conditions (generalized Pauli constraints). Instead, the complexity may simply be transferred from the  underlying space of pure $N$-representable one-matrices to the structure of the universal one-matrix functional.
\end{abstract}

\maketitle

\section{Introduction}
Reduced density matrix functional theory (RDMFT) \cite{G75,C00,M07,PG16,SKB17} extends the widely used  density functional theory (DFT) \cite{HK64,PY95,GD95,J15} by involving the full one-particle reduced density matrix (1RDM) $\gamma$ rather than just the spatial density. This therefore facilitates the exact description of the energy of any one-particle Hamiltonian $h$ (including, e.g., the kinetic energy or a non-local external potential). Furthermore, RDMFT allows explicitly for fractional occupation numbers as it is required in the description of strongly correlated systems \cite{PG16} and thus offers promising prospects of overcoming the fundamental limitations of DFT\cite{LHG07,LM08}. At the same time, involving the full 1RDM leads also to drawbacks relative to DFT:
The complexity of the $N$-representability problem, e.g., is not only hidden in the structure of the universal functional as in DFT\cite{SV09}  but even the space of underlying 1RDMs is already non-trivial. To explain this aspect crucial to our work, we consider Hamiltonians of the form $H=h+V$ on the $N$-fermion Hilbert space $\mathcal{H}_N\equiv\wedge^N[\mathcal{H}_1]$, where $h$ is a one-particle Hamiltonian and $V$ some interaction (e.g.~Coulomb pair interaction) which is fixed for the following. Moreover, we assume a finite-dimensional one-particle Hilbert space $\mathcal{H}_1$ and denote the convex set of $N$-fermion density operators $\Gamma$ by $\mathcal{E}^N$ and the subset of pure states by $\mathcal{P}^N$. A general expression for the universal functional\cite{LE79} follows then immediately by determining the ground state energy of $H$
\begin{eqnarray}\label{Levy}
  E(h) &=& \min_{\Gamma\in \mathcal{P}^N} \mbox{Tr}_N[(h+V)\Gamma] \nonumber \\
  &=& \min_{\gamma\in \mathcal{P}^1_N}\Big[\mbox{Tr}_1[h\gamma]+\min_{\mathcal{P}^N\ni\Gamma\mapsto \gamma}\mbox{Tr}_N[V\Gamma]\Big]\nonumber \\
 &\equiv&  \min_{\gamma\in \mathcal{P}^1_N}\Big[\mbox{Tr}_1[h\gamma]+\mathcal{F}_p(\gamma)\Big]\,.
\end{eqnarray}
In the second line, we have introduced the set $\mathcal{P}^1_N$ of pure $N$-representable 1RDMs $\gamma$ and the last line gives rise to the universal pure functional $\mathcal{F}_p$ defined on $\mathcal{P}^1_N$. The crucial point is now that $\mathcal{P}^1_N$ is not only constrained by the simple Pauli exclusion principle, $0\leq \gamma \leq 1$, but there are rather involved additional one-body \emph{pure} $N$-representability conditions (generalized Pauli constraints), linear conditions on the eigenvalues of the 1RDM\cite{KL06,AK08,KL09}. To circumvent at first sight the complexity of those generalized Pauli constraints, Valone proposed\cite{V80} to relax in \eqref{Levy} the set $\mathcal{P}^N$ to $\mathcal{E}^N$ by skipping the purity, leading to
\begin{equation}\label{Valone}
  E(h)  = \min_{\gamma\in \mathcal{E}^1_N}\Big[\mbox{Tr}_1[h\gamma]+\mathcal{F}_e(\gamma)\Big]
\end{equation}
with the ensemble functional $\mathcal{F}_e(\gamma)\equiv \min_{\mathcal{E}^N\ni\Gamma\mapsto \gamma}\mbox{Tr}_N[V\Gamma]$ defined on the convex set $\mathcal{E}^1_N$ of ensemble $N$-representable 1RDMs.
One may now expect that the complexity of the one-body pure $N$-representability conditions is simply transferred within ensemble RDMFT from the underlying set of 1RDMs to the structure of the exact functional $\mathcal{F}_e$. This, however, seems not to happen according to Ref.~\onlinecite{NLT85}, suggesting and proving that $\mathcal{F}_e$ and $\mathcal{F}_p$ coincide on their common domain $\mathcal{P}^1_N$ of pure $N$-representable 1RDMs, $\mathcal{F}_p\equiv \mathcal{F}_e|_{\mathcal{P}^1_N}$. In our work, we refute this fundamental theorem in RDMFT and show that the ensemble functional follows instead as the lower convex envelop of the pure functional. For this, we first need to develop a better understanding for the space of $N$-fermion density matrices exploited in Levy's constrained search\cite{LE79},
i.e.~the sets
\begin{eqnarray}\label{Nsetsgamma}
\mathcal{P}^N(\gamma)&\equiv& \{\Gamma \in \mathcal{P}^N|\Gamma \mapsto \gamma\}\nonumber \\
\mathcal{E}^N(\gamma) &\equiv& \{\Gamma \in \mathcal{E}^N|\Gamma \mapsto \gamma\}
\end{eqnarray}
of $N$-fermion density operators $\Gamma$ mapping to a given 1RDM $\gamma$.

\section{An instructive example: Hubbard dimer}
The simplest way to refute the suggested equality $\mathcal{F}_p\equiv \mathcal{F}_e|_{\mathcal{P}^1_N}$ is to find one counterexample.
A simple one is given by the \emph{asymmetric} Hubbard dimer,
\begin{eqnarray} \label{hamil}
H&=&  -t \sum_{\sigma}\big[ c^{\dagger}_{1\sigma}c_{2\sigma}  + c^{\dagger}_{2\sigma}c_{1\sigma}\big] + \sum_{\sigma}\big[\epsilon_1 n_{1\sigma} + \epsilon_2 n_{2\sigma} \big] \nonumber \\
&+&U \big[n_{1\uparrow}n_{1\downarrow}  + n_{2\uparrow}n_{2\downarrow}\big]\,,
\end{eqnarray}
a system of two electrons on two sites. Here, $c^{\dagger}_{i\sigma}$($c_{i\sigma}$) denotes the creation(annihilation) operator of an electron at site $i$ with spin $\sigma$ and $n_{i\sigma}=c^{\dagger}_{i\sigma}c_{i\sigma}$ is the corresponding occupation number operator. The first two terms in Eq.~(\ref{hamil}) represents the kinetic and external potential energy, while the last one describes the interaction ($V$) between the electrons. We restrict $H$ to the three-dimensional singlet space which contains the ground state. It is an elementary exercise\cite{SP11,CMS16} to determine the respective pure functional $\mathcal{F}_p$,
\begin{equation} \label{Fp}
 \mathcal{F}_p(\gamma)= U\frac{\frac{1}{2}\gamma_{12}^2   \big[1-\sqrt{1-4\gamma_{12}^2-4(\gamma_{11}-\frac{1}{2})^2}   \big] +(\gamma_{11}-\frac{1}{2})^2}{\gamma_{12}^2+(\gamma_{11}-\frac{1}{2})^2} ,
\end{equation}
with $\gamma_{ij}\equiv \langle i\!\uparrow\!|\gamma|j\!\uparrow\rangle = \langle i\!\downarrow\!|\gamma|j\!\downarrow \rangle$, $i,j=1,2$.
$\mathcal{F}_p(\gamma)$ is invariant under $\gamma_{11} \to (1-\gamma_{11}) $ (particle-hole duality \cite{Y01}) and $\gamma_{12} \to -\gamma_{12}$.
\begin{figure}[htb]
\includegraphics[scale=0.18]{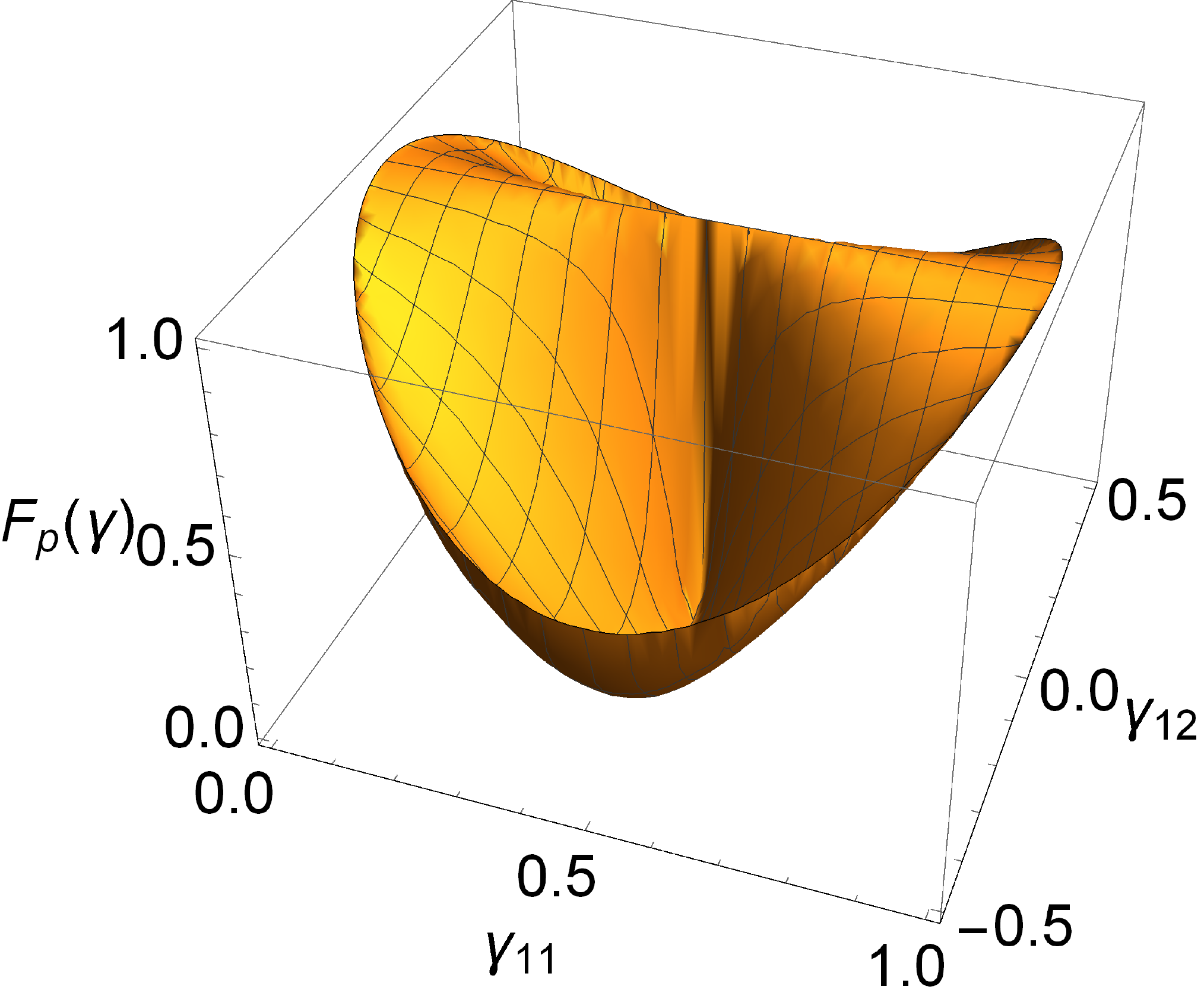}
\includegraphics[scale=0.18]{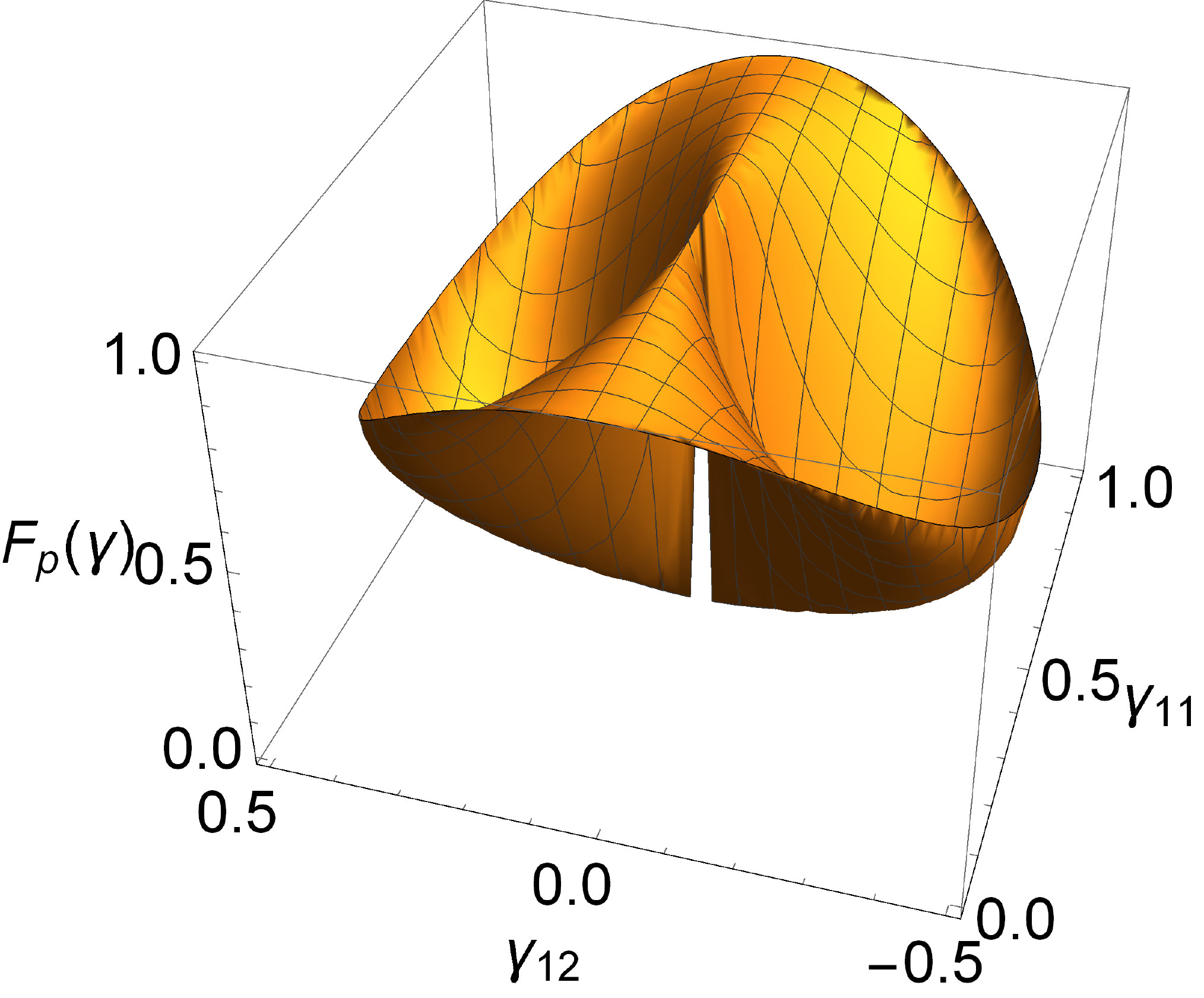}
\caption{$\mathcal{F}_p(\gamma)$ (in units of $U$) as a function of $\gamma_{11}$ and $\gamma_{12}$}
\label{fig:Fpdimer}
\end{figure}

On the one hand, result \eqref{Fp} and its graphical illustration in Fig.~\ref{fig:Fpdimer} reveal that the pure functional $\mathcal{F}_p$ for the Hubbard dimer \eqref{hamil} is not convex on the set $\mathcal{P}^1_2=\mathcal{E}^1_2$ (described by the condition\cite{SP11,CMS16}
$(\gamma_{11}-\frac{1}{2})^2 + \gamma_{12}^2 \leq \frac{1}{4})$). On the other hand, it is well-known\cite{ZM85,PG16} and rather elementary to verify that the ensemble functional $\mathcal{F}_e$ is \emph{always} convex. As a consequence, the Hubbard dimer already refutes the suggested equality $\mathcal{F}_e\equiv \mathcal{F}_p$ on $\mathcal{P}^1_2$.

\section{Geometric picture of Levy's constrained search}
It is instructive to understand the geometric picture of density matrices $\Gamma$ underlying Levy's constrained search \eqref{Levy} and \eqref{Valone}. This will in particular reveal the loophole in the derivation in Ref.~\onlinecite{NLT85}.
Let us first recall that the set $\mathcal{E}^N$ of $N$-fermion density matrices is convex and also compact as a subset of the space of hermitian matrices with fixed trace (i.e.~it is bounded and closed). Its extremal points are given by the pure states, forming the compact but non-convex set $\mathcal{P}^N$. These are the idempotent matrices, $\Gamma=\Gamma^2$ (i.e.~their eigenvalues all vanish except one).
It is worth noticing that a ``point'' $\Gamma$ in $\mathcal{E}^N$ lies on the boundary if and only if $\Gamma$ is not strictly positive, i.e., at least one of its eigenvalues vanishes. As a consequence, most boundary points are not extremal points. It is one of the crucial insights of our work that this changes considerably if we restrict this consideration to the subsets $\mathcal{E}^N(\gamma)$ and $\mathcal{P}^N(\gamma)$ with respect to which the minimization \eqref{Nsetsgamma} is carried out: While both sets $\mathcal{E}^N(\gamma)$ and $\mathcal{P}^N(\gamma)$ are also compact and $\mathcal{E}^N(\gamma)$ is convex for all $\gamma$, extremal states $\Gamma$ of $\mathcal{E}^N(\gamma)$ are not necessarily pure anymore. The general reason for this is that a convex decompositions of $\Gamma$ (e.g.~the spectral decomposition into pure states) involves states whose 1RDMs typically differ from $\gamma$. Thus, a mixed (i.e.~non-pure) $\Gamma$ might be extremal within $\mathcal{E}^N(\gamma)$ despite the fact that it is not extremal within $\mathcal{E}^N$.

As already stated above, the ensemble functional $\mathcal{F}_e(\gamma)$ follows for each $\gamma \in \mathcal{E}^1_N$ by minimizing $\mbox{Tr}_N[V\Gamma]$ over $\mathcal{E}^N(\gamma)$. Since $\mbox{Tr}_N[V(\cdot)]$ is linear and $\mathcal{E}^N(\gamma)$ convex and compact, the minimum (i.e.~$\mathcal{F}_e(\gamma)$) is attained on the boundary of $\mathcal{E}^N(\gamma)$. This is a general (and rather obvious) fact from linear optimization: First, we observe that $\mbox{Tr}_N[V\Gamma]$ is nothing else than the standard inner product on the Hilbert space of hermitian matrices, $\mbox{Tr}_N[V\Gamma]\equiv \langle V,\Gamma\rangle_N$. In that sense, there is given a notion of geometry on the space of density operators\cite{C63,K67,E72,D76,H78a,H78b,H92,PS96,B11,OCZYJRC11,M12,CJRZZ12} and $V$ defines thus a direction in $\mathcal{E}^N(\gamma)$. The set of $\Gamma \in \mathcal{E}^N(\gamma)$ with a specific interaction energy $v=\langle V,\Gamma\rangle_N$ gives rise to a hyperplane, orthogonal to $V$. The minimum of $\mbox{Tr}_N[V(\cdot)]\equiv \langle V,\cdot\rangle_N$ on $\mathcal{E}^N(\gamma)$ then follows by shifting the hyperplane along the direction $-V$ (i.e.~by reducing $v$) until an extremal point of $\mathcal{E}^N(\gamma)$ is reached. This final hyperplane is a so-called \emph{supporting hyperplane}\cite{R97}. By definition, this means that $\mathcal{E}^N(\gamma)$ is entirely contained in one of the two closed half-spaces bounded by that hyperplane and $\mathcal{E}^N(\gamma)$ has at least one boundary point on the hyperplane.
\begin{figure}[htb]
\includegraphics[width=0.68\columnwidth]{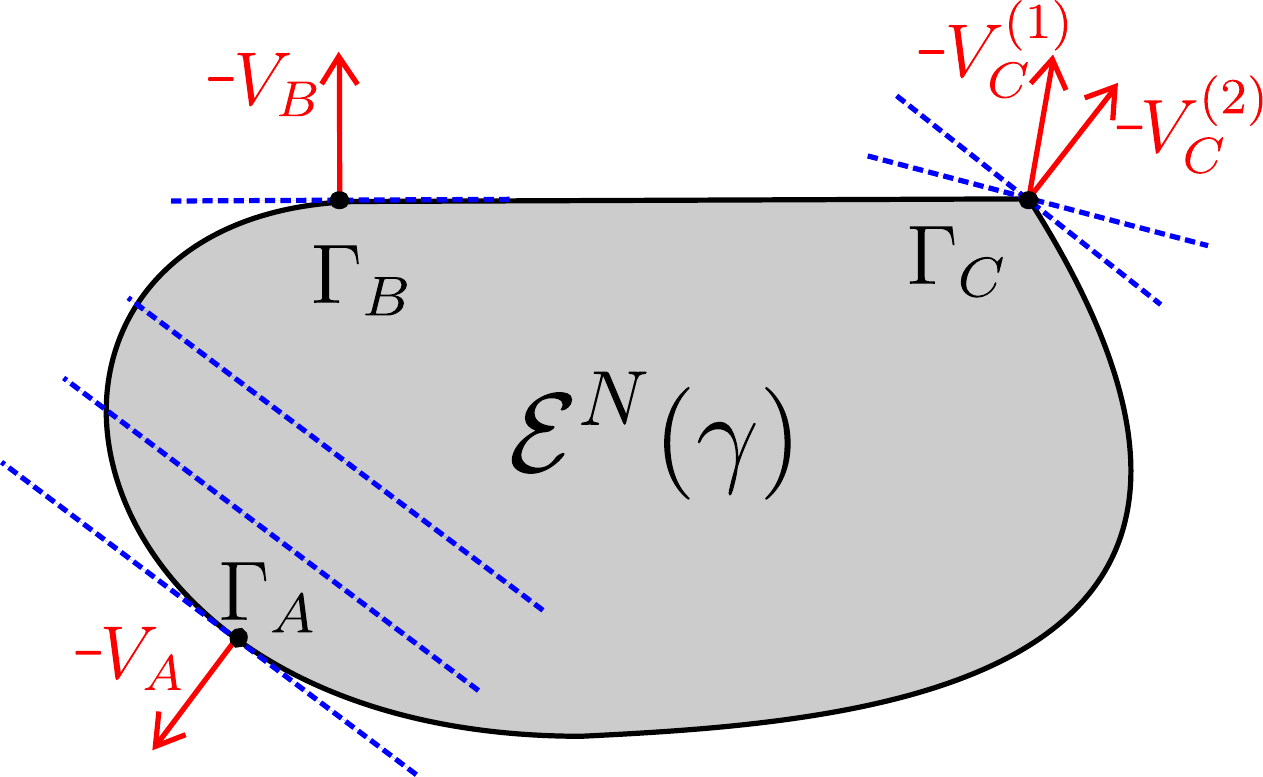}
\caption{Schematic illustration of the geometric picture underlying Levy's constrained search for determining the ensemble functional $\mathcal{F}_e(\gamma)$: For each  1RDM $\gamma$, the linear functional $\mbox{Tr}_N[V(\cdot)]$ attains it minimum (i.e.~$\mathcal{F}_e(\gamma)$) on the boundary of $\mathcal{E}^N(\gamma)$ (see also text).}
\label{fig:extr1}
\end{figure}
This geometric picture underlying Levy's constrained search is illustrated in Fig.~\ref{fig:extr1} for different interactions (``directions'') $V$. There are three conceptually different boundary points which can be characterized by referring to two distinctive features: On the one hand, point $\Gamma_A$ and $\Gamma_B$ have a unique supporting hyperplane (unique ``normal'' vector $V$), in contrast to $\Gamma_C$ supported by infinitely many hyperplanes. On the other hand, point $\Gamma_A$ and $\Gamma_C$ are \emph{exposed}\cite{R97} in contrast to $\Gamma_B$, i.e.~they are supported by hyperplanes which do not contain any further boundary points. In other words, $\Gamma_A$ and $\Gamma_C$ can be obtained as \emph{unique} minimizers of $\mbox{Tr}_N[V\Gamma]$ for some $V$.

After having explained the geometric picture underlying Levy's constrained search, we can now identify the loophole of the proof in Ref.~\onlinecite{NLT85} which we briefly recap: For $\gamma \in \mathcal{P}^1_N$ the minimizer of $\mbox{Tr}_N[V \Gamma]$ on $\mathcal{E}^N(\gamma)$ is denoted by
$\overline{\Gamma}$. Since $\mathcal{E}^N(\gamma)$ is convex and compact, $\overline{\Gamma}$ can be expressed according to the Krein-Milman theorem\cite{KM40} as a convex combination of the extreme points of $\mathcal{E}^N(\gamma)$. This convex combination can be grouped into two parts, one ($w_p \Gamma_p$) arising from pure extremal states and one ($(1-w_p) \Gamma_e$) arising from mixed states, i.e.~$\mathcal{F}_e(\gamma)= \mbox{Tr}_N[V \overline{\Gamma}]=w_p \mbox{Tr}_N[V \Gamma_p]+(1-w_p) \mbox{Tr}_N[V \Gamma_e]$ (with $\Gamma_{p/e}$ normalized to unity). In a straightforward manner\cite{NLT85} this yields $\mathcal{F}_e(\gamma)\geq w_p \mathcal{F}_p(\gamma) + (1-w_p)\mathcal{F}_e(\gamma)$, implying (if $w_p>0$) $\mathcal{F}_e(\gamma)\geq \mathcal{F}_p(\gamma)$. In combination with $\mathcal{F}_e(\gamma)\leq \mathcal{F}_p(\gamma)$ (following from the definition of $\mathcal{F}_{p/e}$ and $\mathcal{P}^N(\gamma)\subset\mathcal{E}^N(\gamma)$), this eventually yields the suggested equality $\mathcal{F}_e(\gamma)= \mathcal{F}_p(\gamma)$ on $\mathcal{P}^1_N$.
It is exactly the hidden assumption $w_p>0$ which is not justified: As explained above, the minimizer $\overline{\Gamma}$ lies already on the boundary of $\mathcal{E}^N(\gamma)$. Even more importantly, according to a theorem from convex optimization\cite{BDL11}, $\overline{\Gamma}$ is with probability one (i.e.~for generic $V$) already extremal and even exposed.
The application of Krein-Milman's theorem is therefore rather meaningless, the assumption $w_p>0$ is violated as long as the minimizer $\overline{\Gamma}$ is not \emph{incidentally} a pure state and thus $\mathcal{F}_e(\gamma)\neq \mathcal{F}_p(\gamma)$. To illustrate all those general aspects we revisit in the following the Hubbard dimer.
\begin{figure}[h]
\includegraphics[width=0.4\columnwidth]{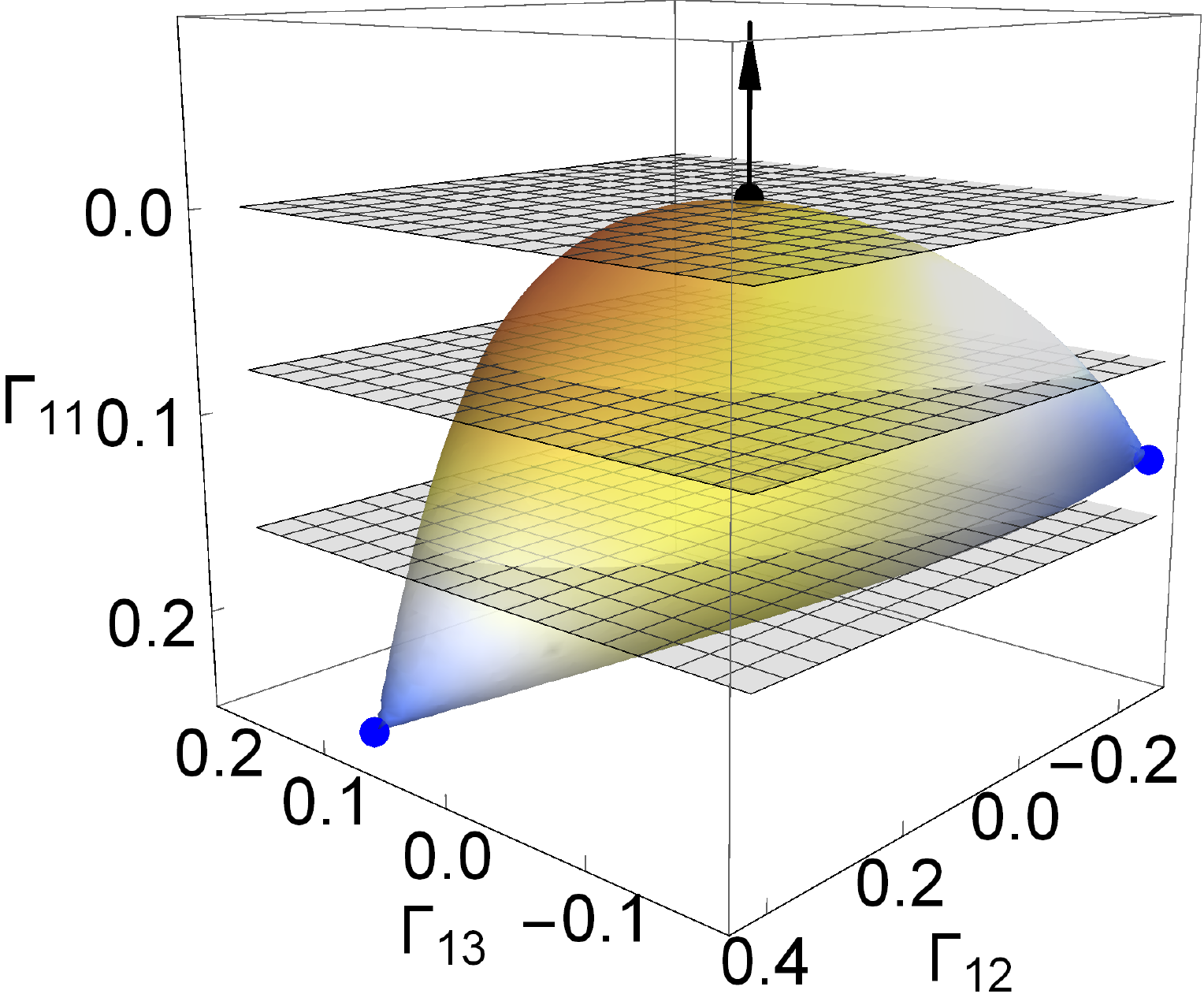}
\includegraphics[width=0.42\columnwidth]{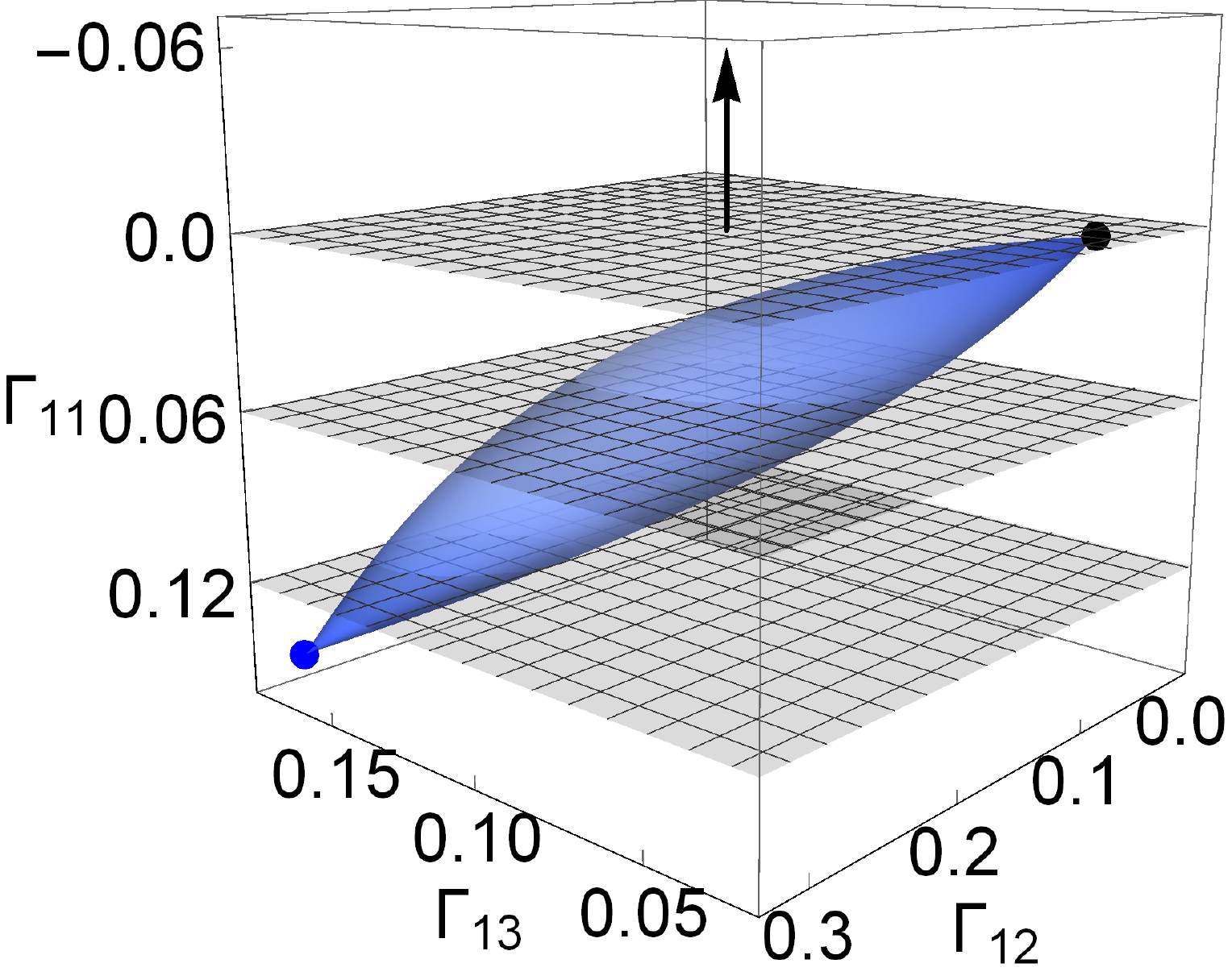}
\includegraphics[width=0.13\columnwidth]{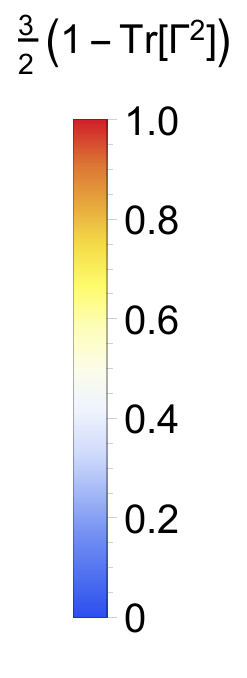}
\caption{Illustration of Levy's constrained search for the Hubbard dimer on $\mathcal{E}^2(\gamma)$ for $(\gamma_{11},\gamma_{12})=(0.25, 0.15)$(left) and $(\gamma_{11},\gamma_{12})=(0.25, 0.38)$(right); the blue dots represent extremal pure $\Gamma$, one of which coincides (right figure) with the minimizer (black dot) of $\mbox{Tr}_2[V\Gamma]$.}
\label{fig:dimerLevy}
\end{figure}

As an orthonormal reference basis for the singlet spin sector underlying the Hubbard dimer we choose $|1\rangle = c^{\dagger}_{1\uparrow}c^{\dagger}_{1\downarrow}|0\rangle$, $|2\rangle = c^{\dagger}_{2\uparrow}c^{\dagger}_{2\downarrow}|0\rangle$ and $|3\rangle =[c^{\dagger}_{1\uparrow}c^{\dagger}_{2\downarrow}|0\rangle -  c^{\dagger}_{1\downarrow}c^{\dagger}_{1\uparrow}|0\rangle ]/\sqrt{2}$, where $|0\rangle$ denotes the vacuum. Expressing any singlet state $\Gamma=\sum_{i,j=1}^3 \Gamma_{ij}\ket{i}\!\bra{j}$ with respect to that basis and restricting it (as usually in quantum chemistry) to real values, the 1RDM in spatial representation follows as (recall $\gamma_{ij}\equiv \langle i\!\uparrow\!|\gamma|j\!\uparrow\rangle = \langle i\!\downarrow\!|\gamma|j\!\downarrow \rangle$, $i,j=1,2$)
\begin{eqnarray} \label{rdm-ensemble}
\gamma_{11}&=&1-\gamma_{22}=\Gamma_{11}+ \frac{1}{2}\Gamma_{33}\nonumber \\
\gamma_{12}&=& \gamma_{21}= \frac{1}{\sqrt{2}}(\Gamma_{13}+\Gamma_{23}) \,.
\end{eqnarray}
The set $\mathcal{E}^2(\gamma)$ can thus be parameterized by three independent real variables. We choose $(\Gamma_{11},\Gamma_{12},\Gamma_{13})$ and find for the expectation value of the Hubbard interaction $\mbox{Tr}_2[V \Gamma]= U (\Gamma_{11}+\Gamma_{22}) = \frac{1}{2} U \big(1+\Gamma_{11}-2\gamma_{11}\big)$ where Eq.~(\ref{rdm-ensemble}) and the normalization of $\Gamma$ have been used.
For two exemplary $\gamma \in \mathcal{P}^1_2\equiv \mathcal{E}^1_2$, we illustrate in Fig.~\ref{fig:dimerLevy} the respective sets $\mathcal{E}^2(\gamma)$. Levy's minimization of the Hubbard interaction $V$ is illustrated as a set of black hyperplanes with the black normal vector corresponding to $-V$. For generic $\gamma$, there are only two pure states on the boundary of $\mathcal{E}^2(\gamma)$, shown as blue dots (in the right figure, one of them is shown in black since it coincides with the minimizer of $\mbox{Tr}_2[V (\cdot)]$). All other points on the boundary turn out to be mixed states (see also color scheme representing the purity $1-\mbox{Tr}[\Gamma^2]$). Since almost all boundary points are exposed and describe mixed states (thus violating the assumption $w_p>0$ in Ref.~\onlinecite{NLT85}) one may now even wonder why the functionals $\mathcal{F}_p$ and $\mathcal{F}_e$ do not differ almost everywhere on $\mathcal{P}^1_2$. The answer to this is the following: By choosing  an interaction $V$ (i.e.~a ``direction'' in $\mathcal{E}^2(\gamma)$) at random, pure states appear with finite probability as minimizers of $\mbox{Tr}_2[V (\cdot)]$. This is due to the fact (see Fig.~\ref{fig:dimerLevy}) that each pure state has a whole range of supporting hyperplanes (see also point $\Gamma_C$ in Fig.~\ref{fig:extr1}), whose normal vectors cover a \emph{finite} angular range.

\section{Relating pure and ensemble functional}
\begin{figure}[htb]
\includegraphics[scale=0.75]{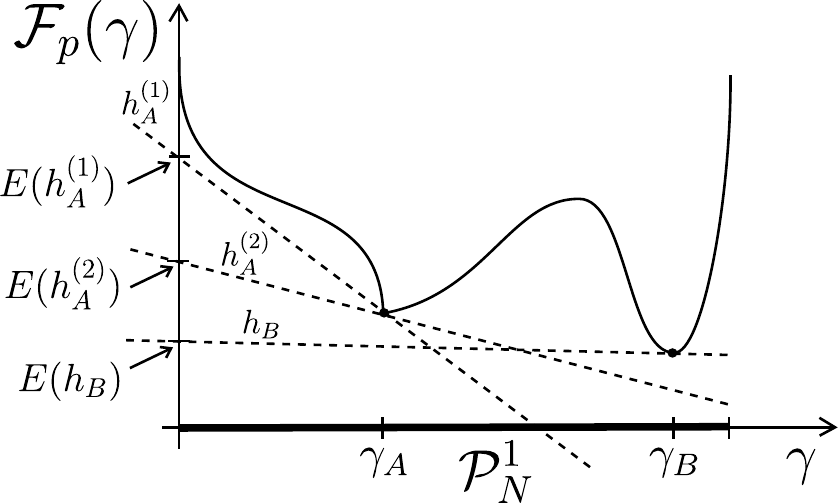}
\caption{Schematic illustration of the energy minimization \eqref{Levy} in RDMFT, emphasizing the role of convex conjugation (Legendre-Fenchel transform) in particular (see also text).}
\label{fig:functional1}
\end{figure}

Strongly inspired by Lieb's seminal work\cite{LI83} on DFT for Coulomb systems, we resort to convex analysis, particularly to the concept of convex conjugation, to relate pure ($\mathcal{F}_p$) and ensemble functional ($\mathcal{F}_e$) for arbitrary interaction $V$. The conjugate $f^*$ (also called Legendre-Fenchel transform) of a function $f: \mathbb{R}^n \rightarrow \mathbb{R}\cup \{\pm \infty\}$ is defined as $f^*(y)=\sup_{x\in \mathbb{R}^n}\big[\langle y,x \rangle-f(x)\big]$. Allowing $f$ to take infinite values ``has the advantage that technical nuisances about effective domains can be suppressed almost entirely''\cite{R97} and we therefore extend $\mathcal{F}_p$ and $\mathcal{F}_e$ to the respective Euclidean space of hermitian matrices by defining $\mathcal{F}_p(\gamma)=\infty$ and $\mathcal{F}_e(\gamma)=\infty$ for $\gamma$ outside their original domains $\mathcal{P}^1_N$ and $\mathcal{E}^1_N$, respectively.
By referring to those common definitions and identifying $\mbox{Tr}_1[h \gamma]$ as the inner product $\langle h,\gamma \rangle$ on the Euclidean space of hermitian matrices, we make the crucial observation that the energy $E(h)$ (recall Eqs.~\eqref{Levy}, \eqref{Valone}) is nothing else than the conjugate of the universal functional $\mathcal{F}_p$ and $\mathcal{F}_e$, respectively (up to an overall minus sign and a reflection $h\mapsto -h$). The conjugation and thus the minimizations in \eqref{Levy} and \eqref{Valone} have a clear geometric meaning as it is illustrated in Fig.~\ref{fig:functional1}: For any fixed `normal vector' $h$, one considers the respective hyperplanes in the Euclidean space of vectors $(\gamma,\mu)$ ($\gamma$ a hermitian matrix, $\mu \in \mathbb{R}$) defined by $\mu = \langle h,\gamma \rangle +u$  and determines the largest $u$ such that the upper closed halfspace of the respective hyperplane still contains the entire graph of $\mathcal{F}_{p/e}$. $E(h)$ then follows as the intercept of that hyperplane with the $\mathcal{F}$-axis, i.e.~the maximal $u$. This interpretation of the conjugation in particular explains in a geometric way why some 1RDMs $\gamma$ (such as those on the line segment between $\gamma_A$ and $\gamma_B$ in Fig.~\ref{fig:functional1}) are not pure $v$-representable\cite{PY95,GD95,NWPSD01,CMS16}.
Moreover, it shows that replacing $\mathcal{F}_p$ in \eqref{Levy} by the \emph{lower convex envelop}\cite{R97}, $\mbox{conv}(\mathcal{F})$, would not change the outcome of the minimization.

\begin{figure}[htb]
\includegraphics[width=0.90\columnwidth]{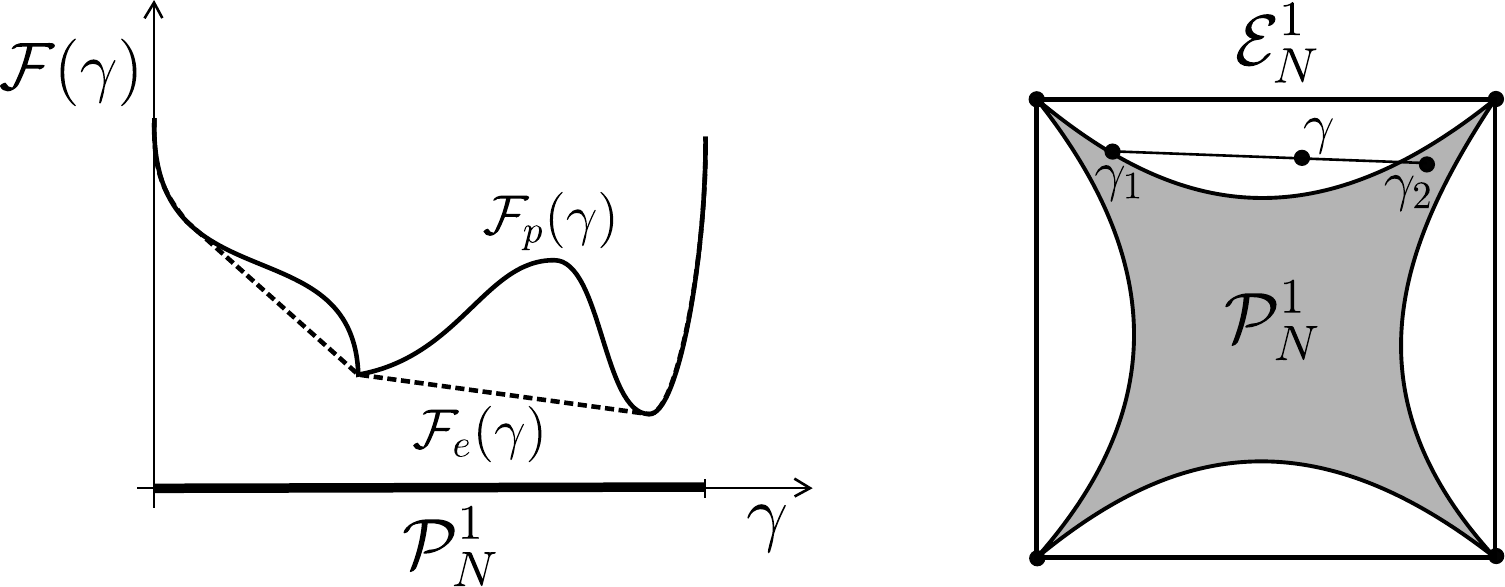}
\caption{Schematic illustration of $\mathcal{F}_e$ given as the lower convex envelop of $\mathcal{F}_p$ (left). This relation between $\mathcal{F}_p$ and $\mathcal{F}_e$  is remarkable since the domain $\mathcal{P}^1_N$ of $\mathcal{F}_p$ is a proper subset of $\mathcal{E}^1_N$ (right), yet their extremal points coincide.}
\label{fig:functional2}
\end{figure}

The second ingredient required for relating $\mathcal{F}_p$ and $\mathcal{F}_e$ is a theorem from convex analysis stating\cite{R97} that
the biconjugate $f^{**}$ coincides with  $f$ whenever $f$ is convex and lower semicontinuous (a weaker form of continuity). Moreover, for arbitrary $f$, $f^{**}$ is (the closure of) the lower convex envelop of $f$. It is straightforward to apply those mathematical results to the functionals $\mathcal{F}_p$, $\mathcal{F}_e$ and $E$: First, since $\mathcal{F}_e$ is convex, it is continuous in the interior
of $\mathcal{E}^1_N$. This implies immediately\cite{R97} lower semicontinuity (except for $\gamma \in \partial \mathcal{E}^1_N$).
We assume in the following  that $\mathcal{F}_e$ is also lower semicontinuous on the boundary $\partial \mathcal{E}^1_N$ of $\mathcal{E}^1_N$. The latter seems to be particularly difficult to verify (also since the interaction $V$ is arbitrary). In case this assumption turns out to be wrong, our final result \eqref{relation} will be valid on the interior of $\mathcal{E}^1_N$ only (which does not reduce at all the significance and scope from any practical point of view). According to the theorem mentioned above, $\mathcal{F}_e$ therefore coincides with its biconjugation. Furthermore, the biconjugate of $\mathcal{F}_p$ coincides with its lower convex envelop $\mbox{conv}(\mathcal{F}_p)$. Yet, since both functionals, $\mathcal{F}_e=\mathcal{F}_e^{**}$ and  $\mbox{conv}(\mathcal{F}_p)=\mathcal{F}_p^{**}$ follow as the conjugate of the same functional, namely the energy $E$ (up to minus signs)
we eventually obtain (see also Fig.~\ref{fig:functional2})
\begin{equation}\label{relation}
\mathcal{F}_e \equiv \mbox{conv}(\mathcal{F}_p)\,.
\end{equation}
It is particularly remarkable that the pure functional $\mathcal{F}_p$ determines the ensemble functional $\mathcal{F}_e$ on its whole domain $\mathcal{E}^1_N$, despite the fact that $\mathcal{F}_p$'s effective domain $\mathcal{P}^1_N$  is a proper subset of $\mathcal{E}^1_N$. To be more specific,  \eqref{relation} namely states that $\mathcal{F}_e(\gamma)$ follows as the minimisation of $\sum_i w_i \mathcal{F}_p(\gamma_i)$ with respect to all possible convex decompositions $\gamma=\sum_i w_i \gamma_i$ ($0\leq w_i\leq1$, $\sum_i w_i=1$) involving only 1RDMs $\gamma_i$ from $\mathcal{P}^1_N$,
$\mathcal{F}_e(\gamma)=\min\!\big\{\sum_i w_i \mathcal{F}_p(\gamma_i)\big|\sum_i w_i \gamma_i=\gamma, \,\gamma_i \in \mathcal{P}^1_N\big\}$.
This is also illustrated on the right panel of Fig.~\ref{fig:functional2} for a $\gamma$ outside $\mathcal{P}^1_N$, also emphasizing the important fact that the extremal points of $\mathcal{P}^1_N$ and $\mathcal{E}^1_N$ coincide.

\section{Summary and conclusion}
A fundamental theorem in RDMFT suggested that the pure ($\mathcal{F}_p$) and ensemble functional ($\mathcal{F}_e$) would coincide on their common domain $\mathcal{P}^1_N$ of pure $N$-representable 1RDMs. Based on a comprehensive study of the geometric picture of density matrices underlying Levy's constrained search, we have refuted this crucial theorem. By exploiting concepts from convex analysis,
we have then shown that $\mathcal{F}_e$ follows instead as the lower convex envelop of $\mathcal{F}_p$. This relation (see Eq.~\eqref{relation}) which holds for any interaction $V$ is particularly remarkable: The pure functional $\mathcal{F}_p$ together with $\mathcal{P}^1_N$ determines the ensemble functional $\mathcal{F}_e$ on its \emph{whole} domain $\mathcal{E}^1_N$, despite the fact that $F_p$'s domain $\mathcal{P}^1_N$ is a proper subset of $\mathcal{E}^1_N$. This letter point in conjunction with the refutation of the relation $\mathcal{F}_p \equiv \mathcal{F}_e|_{\mathcal{P}^1_N}$
demonstrates that relaxing pure RDMFT to ensemble RDMFT does not necessarily circumvent the complexity of the one-body pure $N$-representability conditions. Instead, it may simply be transferred from the underlying space of pure $N$-representable one-matrices into the structure of the universal one-matrix functional $\mathcal{F}_e$. In that case, an additional conceptual insight would follow: Approximating the universal functional would have at least the same computational complexity as the problem of determining all generalized Pauli constraints. Moreover, taking the generalized Pauli constraints into account may facilitate the development of more accurate functionals.

\begin{acknowledgments}
We are very grateful to E.J.\hspace{0.5mm}Baerends, \mbox{O.\hspace{0.5mm}Gritsenko}, \mbox{D.\hspace{0.5mm}Kooi}, \mbox{N.N.\hspace{0.5mm}Lathiotakis}, \mbox{M.\hspace{0.5mm}Piris} and particularly also to \mbox{K.J.H.\hspace{0.5mm}Giesbertz} for inspiring and helpful discussions.
C.S.~acknowledges financial support from the UK Engineering and Physical Sciences Research Council (Grant EP/P007155/1).
\end{acknowledgments}


\begin{thebibliography}{39}%
\makeatletter
\providecommand \@ifxundefined [1]{%
 \@ifx{#1\undefined}
}%
\providecommand \@ifnum [1]{%
 \ifnum #1\expandafter \@firstoftwo
 \else \expandafter \@secondoftwo
 \fi
}%
\providecommand \@ifx [1]{%
 \ifx #1\expandafter \@firstoftwo
 \else \expandafter \@secondoftwo
 \fi
}%
\providecommand \natexlab [1]{#1}%
\providecommand \enquote  [1]{``#1''}%
\providecommand \bibnamefont  [1]{#1}%
\providecommand \bibfnamefont [1]{#1}%
\providecommand \citenamefont [1]{#1}%
\providecommand \href@noop [0]{\@secondoftwo}%
\providecommand \href [0]{\begingroup \@sanitize@url \@href}%
\providecommand \@href[1]{\@@startlink{#1}\@@href}%
\providecommand \@@href[1]{\endgroup#1\@@endlink}%
\providecommand \@sanitize@url [0]{\catcode `\\12\catcode `\$12\catcode
  `\&12\catcode `\#12\catcode `\^12\catcode `\_12\catcode `\%12\relax}%
\providecommand \@@startlink[1]{}%
\providecommand \@@endlink[0]{}%
\providecommand \url  [0]{\begingroup\@sanitize@url \@url }%
\providecommand \@url [1]{\endgroup\@href {#1}{\urlprefix }}%
\providecommand \urlprefix  [0]{URL }%
\providecommand \Eprint [0]{\href }%
\providecommand \doibase [0]{http://dx.doi.org/}%
\providecommand \selectlanguage [0]{\@gobble}%
\providecommand \bibinfo  [0]{\@secondoftwo}%
\providecommand \bibfield  [0]{\@secondoftwo}%
\providecommand \translation [1]{[#1]}%
\providecommand \BibitemOpen [0]{}%
\providecommand \bibitemStop [0]{}%
\providecommand \bibitemNoStop [0]{.\EOS\space}%
\providecommand \EOS [0]{\spacefactor3000\relax}%
\providecommand \BibitemShut  [1]{\csname bibitem#1\endcsname}%
\let\auto@bib@innerbib\@empty
\bibitem [{\citenamefont {Gilbert}(1975)}]{G75}%
  \BibitemOpen
  \bibfield  {author} {\bibinfo {author} {\bibfnamefont {T.~L.}\ \bibnamefont
  {Gilbert}},\ }\href {\doibase 10.1103/PhysRevB.12.2111} {\bibfield  {journal}
  {\bibinfo  {journal} {Phys. Rev. B}\ }\textbf {\bibinfo {volume} {12}},\
  \bibinfo {pages} {2111} (\bibinfo {year} {1975})}\BibitemShut {NoStop}%
\bibitem [{\citenamefont {Cioslowski}(2000)}]{C00}%
  \BibitemOpen
  \bibfield  {author} {\bibinfo {author} {\bibfnamefont {J.}~\bibnamefont
  {Cioslowski}},\ }\href {https://www.springer.com/la/book/9780306464546}
  {\emph {\bibinfo {title} {Many-electron densities and reduced density
  matrices}}}\ (\bibinfo  {publisher} {Springer Science \& Business Media},\
  \bibinfo {year} {2000})\BibitemShut {NoStop}%
\bibitem [{\citenamefont {Piris}(2007)}]{M07}%
  \BibitemOpen
  \bibfield  {author} {\bibinfo {author} {\bibfnamefont {M.}~\bibnamefont
  {Piris}},\ }\enquote {\bibinfo {title} {Natural orbital functional theory},}\
  in\ \href {\doibase 10.1002/9780470106600.ch14} {\emph {\bibinfo {booktitle}
  {Reduced-Density-Matrix Mechanics: With Application to Many-Electron Atoms
  and Molecules}}},\ \bibinfo {editor} {edited by\ \bibinfo {editor}
  {\bibfnamefont {D.~A.}\ \bibnamefont {Mazziotti}}}\ (\bibinfo  {publisher}
  {Wiley-Blackwell},\ \bibinfo {year} {2007})\ Chap.~\bibinfo {chapter} {14},
  p.\ \bibinfo {pages} {387}\BibitemShut {NoStop}%
\bibitem [{\citenamefont {Pernal}\ and\ \citenamefont
  {Giesbertz}(2016)}]{PG16}%
  \BibitemOpen
  \bibfield  {author} {\bibinfo {author} {\bibfnamefont {K.}~\bibnamefont
  {Pernal}}\ and\ \bibinfo {author} {\bibfnamefont {K.~J.~H.}\ \bibnamefont
  {Giesbertz}},\ }\enquote {\bibinfo {title} {Reduced density matrix functional
  theory ({RDMFT}) and linear response time-dependent rdmft ({TD}-{RDMFT})},}\
  in\ \href {\doibase 10.1007/128_2015_624} {\emph {\bibinfo {booktitle}
  {Density-Functional Methods for Excited States}}},\ \bibinfo {editor} {edited
  by\ \bibinfo {editor} {\bibfnamefont {N.}~\bibnamefont {Ferr{\'e}}}, \bibinfo
  {editor} {\bibfnamefont {M.}~\bibnamefont {Filatov}}, \ and\ \bibinfo
  {editor} {\bibfnamefont {M.}~\bibnamefont {Huix-Rotllant}}}\ (\bibinfo
  {publisher} {Springer International Publishing},\ \bibinfo {address} {Cham},\
  \bibinfo {year} {2016})\ p.\ \bibinfo {pages} {125}\BibitemShut {NoStop}%
\bibitem [{\citenamefont {Schade}, \citenamefont {Kamil},\ and\ \citenamefont
  {Bl{\"o}chl}(2017)}]{SKB17}%
  \BibitemOpen
  \bibfield  {author} {\bibinfo {author} {\bibfnamefont {R.}~\bibnamefont
  {Schade}}, \bibinfo {author} {\bibfnamefont {E.}~\bibnamefont {Kamil}}, \
  and\ \bibinfo {author} {\bibfnamefont {P.}~\bibnamefont {Bl{\"o}chl}},\
  }\href {\doibase 10.1140/epjst/e2017-70046-0} {\bibfield  {journal} {\bibinfo
   {journal} {Eur. Phys. J. Special Topics}\ }\textbf {\bibinfo {volume}
  {226}},\ \bibinfo {pages} {2677} (\bibinfo {year} {2017})}\BibitemShut
  {NoStop}%
\bibitem [{\citenamefont {Hohenberg}\ and\ \citenamefont {Kohn}(1964)}]{HK64}%
  \BibitemOpen
  \bibfield  {author} {\bibinfo {author} {\bibfnamefont {P.}~\bibnamefont
  {Hohenberg}}\ and\ \bibinfo {author} {\bibfnamefont {W.}~\bibnamefont
  {Kohn}},\ }\href {\doibase 10.1103/PhysRev.136.B864} {\bibfield  {journal}
  {\bibinfo  {journal} {Phys. Rev.}\ }\textbf {\bibinfo {volume} {136}},\
  \bibinfo {pages} {B864} (\bibinfo {year} {1964})}\BibitemShut {NoStop}%
\bibitem [{\citenamefont {Parr}\ and\ \citenamefont {Yang}(1995)}]{PY95}%
  \BibitemOpen
  \bibfield  {author} {\bibinfo {author} {\bibfnamefont {R.~G.}\ \bibnamefont
  {Parr}}\ and\ \bibinfo {author} {\bibfnamefont {W.}~\bibnamefont {Yang}},\
  }\href
  {https://www.annualreviews.org/doi/abs/10.1146/annurev.pc.46.100195.003413}
  {\bibfield  {journal} {\bibinfo  {journal} {Annu. Rev. Phys. Chem.}\ }\textbf
  {\bibinfo {volume} {46}},\ \bibinfo {pages} {701} (\bibinfo {year}
  {1995})}\BibitemShut {NoStop}%
\bibitem [{\citenamefont {Gross}\ and\ \citenamefont {Dreizler}(2013)}]{GD95}%
  \BibitemOpen
  \bibfield  {author} {\bibinfo {author} {\bibfnamefont {E.}~\bibnamefont
  {Gross}}\ and\ \bibinfo {author} {\bibfnamefont {R.}~\bibnamefont
  {Dreizler}},\ }\href {https://www.springer.com/la/book/9781475799774} {\emph
  {\bibinfo {title} {Density functional theory}}},\ Vol.\ \bibinfo {volume}
  {337}\ (\bibinfo  {publisher} {Springer Science \& Business Media},\ \bibinfo
  {year} {2013})\BibitemShut {NoStop}%
\bibitem [{\citenamefont {Jones}(2015)}]{J15}%
  \BibitemOpen
  \bibfield  {author} {\bibinfo {author} {\bibfnamefont {R.~O.}\ \bibnamefont
  {Jones}},\ }\href {\doibase 10.1103/RevModPhys.87.897} {\bibfield  {journal}
  {\bibinfo  {journal} {Rev. Mod. Phys.}\ }\textbf {\bibinfo {volume} {87}},\
  \bibinfo {pages} {897} (\bibinfo {year} {2015})}\BibitemShut {NoStop}%
\bibitem [{\citenamefont {Lathiotakis}, \citenamefont {Helbig},\ and\
  \citenamefont {Gross}(2007)}]{LHG07}%
  \BibitemOpen
  \bibfield  {author} {\bibinfo {author} {\bibfnamefont {N.~N.}\ \bibnamefont
  {Lathiotakis}}, \bibinfo {author} {\bibfnamefont {N.}~\bibnamefont {Helbig}},
  \ and\ \bibinfo {author} {\bibfnamefont {E.~K.~U.}\ \bibnamefont {Gross}},\
  }\href {\doibase 10.1103/PhysRevB.75.195120} {\bibfield  {journal} {\bibinfo
  {journal} {Phys. Rev. B}\ }\textbf {\bibinfo {volume} {75}},\ \bibinfo
  {pages} {195120} (\bibinfo {year} {2007})}\BibitemShut {NoStop}%
\bibitem [{\citenamefont {Lathiotakis}\ and\ \citenamefont
  {Marques}(2008)}]{LM08}%
  \BibitemOpen
  \bibfield  {author} {\bibinfo {author} {\bibfnamefont {N.~N.}\ \bibnamefont
  {Lathiotakis}}\ and\ \bibinfo {author} {\bibfnamefont {M.~A.~L.}\
  \bibnamefont {Marques}},\ }\href
  {https://aip.scitation.org/doi/abs/10.1063/1.2899328} {\bibfield  {journal}
  {\bibinfo  {journal} {J. Chem. Phys.}\ }\textbf {\bibinfo {volume} {128}},\
  \bibinfo {pages} {184103} (\bibinfo {year} {2008})}\BibitemShut {NoStop}%
\bibitem [{\citenamefont {Schuch}\ and\ \citenamefont
  {Verstraete}(2009)}]{SV09}%
  \BibitemOpen
  \bibfield  {author} {\bibinfo {author} {\bibfnamefont {N.}~\bibnamefont
  {Schuch}}\ and\ \bibinfo {author} {\bibfnamefont {F.}~\bibnamefont
  {Verstraete}},\ }\href {https://www.nature.com/articles/nphys1370} {\bibfield
   {journal} {\bibinfo  {journal} {Nat. Phys.}\ }\textbf {\bibinfo {volume}
  {5}},\ \bibinfo {pages} {732} (\bibinfo {year} {2009})}\BibitemShut {NoStop}%
\bibitem [{\citenamefont {Levy}(1979)}]{LE79}%
  \BibitemOpen
  \bibfield  {author} {\bibinfo {author} {\bibfnamefont {M.}~\bibnamefont
  {Levy}},\ }\href {http://www.pnas.org/content/76/12/6062} {\bibfield
  {journal} {\bibinfo  {journal} {Proc. Natl. Acad. Sci. U.S.A}\ }\textbf
  {\bibinfo {volume} {76}},\ \bibinfo {pages} {6062} (\bibinfo {year}
  {1979})}\BibitemShut {NoStop}%
\bibitem [{\citenamefont {Klyachko}(2006)}]{KL06}%
  \BibitemOpen
  \bibfield  {author} {\bibinfo {author} {\bibfnamefont {A.}~\bibnamefont
  {Klyachko}},\ }\href {http://stacks.iop.org/1742-6596/36/i=1/a=014}
  {\bibfield  {journal} {\bibinfo  {journal} {J. Phys. Conf. Ser.}\ }\textbf
  {\bibinfo {volume} {36}},\ \bibinfo {pages} {72} (\bibinfo {year}
  {2006})}\BibitemShut {NoStop}%
\bibitem [{\citenamefont {Altunbulak}\ and\ \citenamefont
  {Klyachko}(2008)}]{AK08}%
  \BibitemOpen
  \bibfield  {author} {\bibinfo {author} {\bibfnamefont {M.}~\bibnamefont
  {Altunbulak}}\ and\ \bibinfo {author} {\bibfnamefont {A.}~\bibnamefont
  {Klyachko}},\ }\href {\doibase 10.1007/s00220-008-0552-z} {\bibfield
  {journal} {\bibinfo  {journal} {Commun. Math. Phys.}\ }\textbf {\bibinfo
  {volume} {282}},\ \bibinfo {pages} {287} (\bibinfo {year}
  {2008})}\BibitemShut {NoStop}%
\bibitem [{\citenamefont {Klyachko}(2009)}]{KL09}%
  \BibitemOpen
  \bibfield  {author} {\bibinfo {author} {\bibfnamefont {A.}~\bibnamefont
  {Klyachko}},\ }\href {http://arxiv.org/abs/0904.2009} {\bibfield  {journal}
  {\bibinfo  {journal} {arXiv:0904.2009}\ } (\bibinfo {year}
  {2009})}\BibitemShut {NoStop}%
\bibitem [{\citenamefont {Valone}(1980)}]{V80}%
  \BibitemOpen
  \bibfield  {author} {\bibinfo {author} {\bibfnamefont {S.~M.}\ \bibnamefont
  {Valone}},\ }\href {\doibase 10.1063/1.440249} {\bibfield  {journal}
  {\bibinfo  {journal} {J. Chem. Phys.}\ }\textbf {\bibinfo {volume} {73}},\
  \bibinfo {pages} {1344} (\bibinfo {year} {1980})}\BibitemShut {NoStop}%
\bibitem [{\citenamefont {Nguyen-Dang}, \citenamefont {Ludena},\ and\
  \citenamefont {Tal}(1985)}]{NLT85}%
  \BibitemOpen
  \bibfield  {author} {\bibinfo {author} {\bibfnamefont {T.~T.}\ \bibnamefont
  {Nguyen-Dang}}, \bibinfo {author} {\bibfnamefont {E.~V.}\ \bibnamefont
  {Ludena}}, \ and\ \bibinfo {author} {\bibfnamefont {Y.}~\bibnamefont {Tal}},\
  }\href {\doibase https://doi.org/10.1016/0166-1280(85)85114-9} {\bibfield
  {journal} {\bibinfo  {journal} {Comput. Theor. Chem.}\ }\textbf {\bibinfo
  {volume} {120}},\ \bibinfo {pages} {247} (\bibinfo {year}
  {1985})}\BibitemShut {NoStop}%
\bibitem [{\citenamefont {Sauban\`ere}\ and\ \citenamefont
  {Pastor}(2011)}]{SP11}%
  \BibitemOpen
  \bibfield  {author} {\bibinfo {author} {\bibfnamefont {M.}~\bibnamefont
  {Sauban\`ere}}\ and\ \bibinfo {author} {\bibfnamefont {G.~M.}\ \bibnamefont
  {Pastor}},\ }\href {\doibase 10.1103/PhysRevB.84.035111} {\bibfield
  {journal} {\bibinfo  {journal} {Phys. Rev. B}\ }\textbf {\bibinfo {volume}
  {84}},\ \bibinfo {pages} {035111} (\bibinfo {year} {2011})}\BibitemShut
  {NoStop}%
\bibitem [{\citenamefont {Cohen}\ and\ \citenamefont
  {Mori-S\'anchez}(2016)}]{CMS16}%
  \BibitemOpen
  \bibfield  {author} {\bibinfo {author} {\bibfnamefont {A.~J.}\ \bibnamefont
  {Cohen}}\ and\ \bibinfo {author} {\bibfnamefont {P.}~\bibnamefont
  {Mori-S\'anchez}},\ }\href {\doibase 10.1103/PhysRevA.93.042511} {\bibfield
  {journal} {\bibinfo  {journal} {Phys. Rev. A}\ }\textbf {\bibinfo {volume}
  {93}},\ \bibinfo {pages} {042511} (\bibinfo {year} {2016})}\BibitemShut
  {NoStop}%
\bibitem [{\citenamefont {Yasuda}(2001)}]{Y01}%
  \BibitemOpen
  \bibfield  {author} {\bibinfo {author} {\bibfnamefont {K.}~\bibnamefont
  {Yasuda}},\ }\href {\doibase 10.1103/PhysRevA.63.032517} {\bibfield
  {journal} {\bibinfo  {journal} {Phys. Rev. A}\ }\textbf {\bibinfo {volume}
  {63}},\ \bibinfo {pages} {032517} (\bibinfo {year} {2001})}\BibitemShut
  {NoStop}%
\bibitem [{\citenamefont {Zumbach}\ and\ \citenamefont {Maschke}(1985)}]{ZM85}%
  \BibitemOpen
  \bibfield  {author} {\bibinfo {author} {\bibfnamefont {G.}~\bibnamefont
  {Zumbach}}\ and\ \bibinfo {author} {\bibfnamefont {K.}~\bibnamefont
  {Maschke}},\ }\href {\doibase 10.1063/1.448595} {\bibfield  {journal}
  {\bibinfo  {journal} {J. Chem. Phys.}\ }\textbf {\bibinfo {volume} {82}},\
  \bibinfo {pages} {5604} (\bibinfo {year} {1985})}\BibitemShut {NoStop}%
\bibitem [{\citenamefont {Coleman}(1963)}]{C63}%
  \BibitemOpen
  \bibfield  {author} {\bibinfo {author} {\bibfnamefont {A.~J.}\ \bibnamefont
  {Coleman}},\ }\href {\doibase 10.1103/RevModPhys.35.668} {\bibfield
  {journal} {\bibinfo  {journal} {Rev. Mod. Phys.}\ }\textbf {\bibinfo {volume}
  {35}},\ \bibinfo {pages} {668} (\bibinfo {year} {1963})}\BibitemShut
  {NoStop}%
\bibitem [{\citenamefont {Kummer}(1967)}]{K67}%
  \BibitemOpen
  \bibfield  {author} {\bibinfo {author} {\bibfnamefont {H.}~\bibnamefont
  {Kummer}},\ }\href {https://aip.scitation.org/doi/10.1063/1.1705122}
  {\bibfield  {journal} {\bibinfo  {journal} {J. Math. Phys.}\ }\textbf
  {\bibinfo {volume} {8}},\ \bibinfo {pages} {2063} (\bibinfo {year}
  {1967})}\BibitemShut {NoStop}%
\bibitem [{\citenamefont {Erdahl}(1972)}]{E72}%
  \BibitemOpen
  \bibfield  {author} {\bibinfo {author} {\bibfnamefont {R.~M.}\ \bibnamefont
  {Erdahl}},\ }\href {https://aip.scitation.org/doi/10.1063/1.1665885}
  {\bibfield  {journal} {\bibinfo  {journal} {J. Math. Phys.}\ }\textbf
  {\bibinfo {volume} {13}},\ \bibinfo {pages} {1608} (\bibinfo {year}
  {1972})}\BibitemShut {NoStop}%
\bibitem [{\citenamefont {Davidson}(2012)}]{D76}%
  \BibitemOpen
  \bibfield  {author} {\bibinfo {author} {\bibfnamefont {E.}~\bibnamefont
  {Davidson}},\ }\href
  {https://www.elsevier.com/books/reduced-density-matrices-in-quantum-chemistry/davidson/978-0-12-205850-9}
  {\emph {\bibinfo {title} {Reduced density matrices in quantum chemistry}}},\
  Vol.~\bibinfo {volume} {6}\ (\bibinfo  {publisher} {Elsevier},\ \bibinfo
  {year} {2012})\BibitemShut {NoStop}%
\bibitem [{\citenamefont {Harriman}(1978{\natexlab{a}})}]{H78a}%
  \BibitemOpen
  \bibfield  {author} {\bibinfo {author} {\bibfnamefont {J.~E.}\ \bibnamefont
  {Harriman}},\ }\href {\doibase 10.1103/PhysRevA.17.1249} {\bibfield
  {journal} {\bibinfo  {journal} {Phys. Rev. A}\ }\textbf {\bibinfo {volume}
  {17}},\ \bibinfo {pages} {1249} (\bibinfo {year}
  {1978}{\natexlab{a}})}\BibitemShut {NoStop}%
\bibitem [{\citenamefont {Harriman}(1978{\natexlab{b}})}]{H78b}%
  \BibitemOpen
  \bibfield  {author} {\bibinfo {author} {\bibfnamefont {J.~E.}\ \bibnamefont
  {Harriman}},\ }\href {\doibase 10.1103/PhysRevA.17.1257} {\bibfield
  {journal} {\bibinfo  {journal} {Phys. Rev. A}\ }\textbf {\bibinfo {volume}
  {17}},\ \bibinfo {pages} {1257} (\bibinfo {year}
  {1978}{\natexlab{b}})}\BibitemShut {NoStop}%
\bibitem [{\citenamefont {H{\"u}bner}(1992)}]{H92}%
  \BibitemOpen
  \bibfield  {author} {\bibinfo {author} {\bibfnamefont {M.}~\bibnamefont
  {H{\"u}bner}},\ }\href
  {https://www.sciencedirect.com/science/article/pii/037596019291004B}
  {\bibfield  {journal} {\bibinfo  {journal} {Phys. Lett. A}\ }\textbf
  {\bibinfo {volume} {163}},\ \bibinfo {pages} {239} (\bibinfo {year}
  {1992})}\BibitemShut {NoStop}%
\bibitem [{\citenamefont {Petz}\ and\ \citenamefont {Sud{\'a}r}(1996)}]{PS96}%
  \BibitemOpen
  \bibfield  {author} {\bibinfo {author} {\bibfnamefont {D.}~\bibnamefont
  {Petz}}\ and\ \bibinfo {author} {\bibfnamefont {C.}~\bibnamefont
  {Sud{\'a}r}},\ }\href {https://aip.scitation.org/doi/10.1063/1.531535}
  {\bibfield  {journal} {\bibinfo  {journal} {J. Math. Phys.}\ }\textbf
  {\bibinfo {volume} {37}},\ \bibinfo {pages} {2662} (\bibinfo {year}
  {1996})}\BibitemShut {NoStop}%
\bibitem [{\citenamefont {Brody}(2011)}]{B11}%
  \BibitemOpen
  \bibfield  {author} {\bibinfo {author} {\bibfnamefont {D.~C.}\ \bibnamefont
  {Brody}},\ }\href
  {http://iopscience.iop.org/article/10.1088/1751-8113/44/25/252002/meta}
  {\bibfield  {journal} {\bibinfo  {journal} {J. Phys. A}\ }\textbf {\bibinfo
  {volume} {44}},\ \bibinfo {pages} {252002} (\bibinfo {year}
  {2011})}\BibitemShut {NoStop}%
\bibitem [{\citenamefont {Ocko}\ \emph {et~al.}(2011)\citenamefont {Ocko},
  \citenamefont {Chen}, \citenamefont {Zeng}, \citenamefont {Yoshida},
  \citenamefont {Ji}, \citenamefont {Ruskai},\ and\ \citenamefont
  {Chuang}}]{OCZYJRC11}%
  \BibitemOpen
  \bibfield  {author} {\bibinfo {author} {\bibfnamefont {S.~A.}\ \bibnamefont
  {Ocko}}, \bibinfo {author} {\bibfnamefont {X.}~\bibnamefont {Chen}}, \bibinfo
  {author} {\bibfnamefont {B.}~\bibnamefont {Zeng}}, \bibinfo {author}
  {\bibfnamefont {B.}~\bibnamefont {Yoshida}}, \bibinfo {author} {\bibfnamefont
  {Z.}~\bibnamefont {Ji}}, \bibinfo {author} {\bibfnamefont {M.~B.}\
  \bibnamefont {Ruskai}}, \ and\ \bibinfo {author} {\bibfnamefont {I.~L.}\
  \bibnamefont {Chuang}},\ }\href {\doibase 10.1103/PhysRevLett.106.110501}
  {\bibfield  {journal} {\bibinfo  {journal} {Phys. Rev. Lett.}\ }\textbf
  {\bibinfo {volume} {106}},\ \bibinfo {pages} {110501} (\bibinfo {year}
  {2011})}\BibitemShut {NoStop}%
\bibitem [{\citenamefont {Mazziotti}(2012)}]{M12}%
  \BibitemOpen
  \bibfield  {author} {\bibinfo {author} {\bibfnamefont {D.~A.}\ \bibnamefont
  {Mazziotti}},\ }\href {\doibase 10.1103/PhysRevLett.108.263002} {\bibfield
  {journal} {\bibinfo  {journal} {Phys. Rev. Lett.}\ }\textbf {\bibinfo
  {volume} {108}},\ \bibinfo {pages} {263002} (\bibinfo {year}
  {2012})}\BibitemShut {NoStop}%
\bibitem [{\citenamefont {Chen}\ \emph {et~al.}(2012)\citenamefont {Chen},
  \citenamefont {Ji}, \citenamefont {Ruskai}, \citenamefont {Zeng},\ and\
  \citenamefont {Zhou}}]{CJRZZ12}%
  \BibitemOpen
  \bibfield  {author} {\bibinfo {author} {\bibfnamefont {J.}~\bibnamefont
  {Chen}}, \bibinfo {author} {\bibfnamefont {Z.}~\bibnamefont {Ji}}, \bibinfo
  {author} {\bibfnamefont {M.~B.}\ \bibnamefont {Ruskai}}, \bibinfo {author}
  {\bibfnamefont {B.}~\bibnamefont {Zeng}}, \ and\ \bibinfo {author}
  {\bibfnamefont {D.-L.}\ \bibnamefont {Zhou}},\ }\href
  {https://aip.scitation.org/doi/10.1063/1.4736842} {\bibfield  {journal}
  {\bibinfo  {journal} {J. Math. Phys.}\ }\textbf {\bibinfo {volume} {53}},\
  \bibinfo {pages} {072203} (\bibinfo {year} {2012})}\BibitemShut {NoStop}%
\bibitem [{\citenamefont {Rockafellar}(1997)}]{R97}%
  \BibitemOpen
  \bibfield  {author} {\bibinfo {author} {\bibfnamefont {R.~T.}\ \bibnamefont
  {Rockafellar}},\ }\href {https://press.princeton.edu/titles/1815.html} {\emph
  {\bibinfo {title} {Convex {A}nalysis}}}\ (\bibinfo  {publisher} {Princeton
  university press},\ \bibinfo {year} {1997})\BibitemShut {NoStop}%
\bibitem [{\citenamefont {Krein}\ and\ \citenamefont {Milman}(1940)}]{KM40}%
  \BibitemOpen
  \bibfield  {author} {\bibinfo {author} {\bibfnamefont {M.}~\bibnamefont
  {Krein}}\ and\ \bibinfo {author} {\bibfnamefont {D.}~\bibnamefont {Milman}},\
  }\href {https://eudml.org/doc/219061} {\bibfield  {journal} {\bibinfo
  {journal} {Stud. Math.}\ }\textbf {\bibinfo {volume} {9}},\ \bibinfo {pages}
  {133} (\bibinfo {year} {1940})}\BibitemShut {NoStop}%
\bibitem [{\citenamefont {Bolte}, \citenamefont {Daniilidis},\ and\
  \citenamefont {Lewis}(2011)}]{BDL11}%
  \BibitemOpen
  \bibfield  {author} {\bibinfo {author} {\bibfnamefont {J.}~\bibnamefont
  {Bolte}}, \bibinfo {author} {\bibfnamefont {A.}~\bibnamefont {Daniilidis}}, \
  and\ \bibinfo {author} {\bibfnamefont {A.~S.}\ \bibnamefont {Lewis}},\ }\href
  {https://pubsonline.informs.org/doi/abs/10.1287/moor.1110.0481} {\bibfield
  {journal} {\bibinfo  {journal} {Math. Oper. Res.}\ }\textbf {\bibinfo
  {volume} {36}},\ \bibinfo {pages} {55} (\bibinfo {year} {2011})}\BibitemShut
  {NoStop}%
\bibitem [{\citenamefont {Lieb}(1983)}]{LI83}%
  \BibitemOpen
  \bibfield  {author} {\bibinfo {author} {\bibfnamefont {E.~H.}\ \bibnamefont
  {Lieb}},\ }\href {\doibase 10.1002/qua.560240302} {\bibfield  {journal}
  {\bibinfo  {journal} {Int. J. Quantum Chem.}\ }\textbf {\bibinfo {volume}
  {24}},\ \bibinfo {pages} {243} (\bibinfo {year} {1983})}\BibitemShut
  {NoStop}%
\bibitem [{\citenamefont {Van~Neck}\ \emph {et~al.}(2001)\citenamefont
  {Van~Neck}, \citenamefont {Waroquier}, \citenamefont {Peirs}, \citenamefont
  {Van~Speybroeck},\ and\ \citenamefont {Dewulf}}]{NWPSD01}%
  \BibitemOpen
  \bibfield  {author} {\bibinfo {author} {\bibfnamefont {D.}~\bibnamefont
  {Van~Neck}}, \bibinfo {author} {\bibfnamefont {M.}~\bibnamefont {Waroquier}},
  \bibinfo {author} {\bibfnamefont {K.}~\bibnamefont {Peirs}}, \bibinfo
  {author} {\bibfnamefont {V.}~\bibnamefont {Van~Speybroeck}}, \ and\ \bibinfo
  {author} {\bibfnamefont {Y.}~\bibnamefont {Dewulf}},\ }\href {\doibase
  10.1103/PhysRevA.64.042512} {\bibfield  {journal} {\bibinfo  {journal} {Phys.
  Rev. A}\ }\textbf {\bibinfo {volume} {64}},\ \bibinfo {pages} {042512}
  (\bibinfo {year} {2001})}\BibitemShut {NoStop}%
\end{thebibliography}
\end{document}